\documentclass[journal=jpclcd,manuscript=article]{achemso}

\usepackage{chemformula} 
\usepackage[T1]{fontenc} 

\author{Mouhui Yan}
\affiliation[SHU]
{Department of Physics, Shanghai University, 200444 Shanghai, China}
\author{Yichen Jin}
\affiliation[SHU]
{Department of Physics, Shanghai University, 200444 Shanghai, China}
\author{Xiaofei Hou}
\affiliation[STech]
{\mbox{School of Physical Science and Technology, ShanghaiTech University, 201210 Shanghai, China}}
\author{Yanfeng Guo}
\affiliation[STech]
{\mbox{School of Physical Science and Technology, ShanghaiTech University, 201210 Shanghai, China}}
\author{Arshak Tsaturyan}
\affiliation[SFedU]
{Institute of Physical and Organic Chemistry, Southern Federal University, \newline 344090 Rostov on Don, Russia}
\author{Anna Makarova}
\affiliation[FU]
{Institut f\"ur Chemie und Biochemie, Freie Universit\"at Berlin, Arnimallee 22, \newline 14195 Berlin, Germany}
\author{Dmitry Smirnov}
\affiliation[TUD]
{Institut f\"ur Festk\"orper- und Materialphysik, Technische Universit\"at Dresden, \newline 01069 Dresden, Germany}
\author{Yuriy Dedkov} 
\affiliation[E3]
{\mbox{Centre of Excellence ENSEMBLE3 Sp. z o.\,o., ul. Wolczynska 133, 01-919 Warsaw, Poland}}
\email{yuriy.dedkov@icloud.com}
\alsoaffiliation[SHU]
{Department of Physics, Shanghai University, 200444 Shanghai, China}
\author{Elena Voloshina}
\affiliation[E3]
{\mbox{Centre of Excellence ENSEMBLE3 Sp. z o.\,o., ul. Wolczynska 133, 01-919 Warsaw, Poland}}
\email{elena.voloshina@icloud.com}
\alsoaffiliation[FU]
{Institut f\"ur Chemie und Biochemie, Freie Universit\"at Berlin, Arnimallee 22, \newline 14195 Berlin, Germany}
\alsoaffiliation[SHU]
{Department of Physics, Shanghai University, 200444 Shanghai, China}

\title[]
{Topological Quasi-2D Semimetal Co$_3$Sn$_2$S$_2$: Insights To Electronic Structure From NEXAFS and Resonant Photoelectron Spectroscopy}

\begin{document}


\begin{tocentry}

\includegraphics[width=\textwidth]{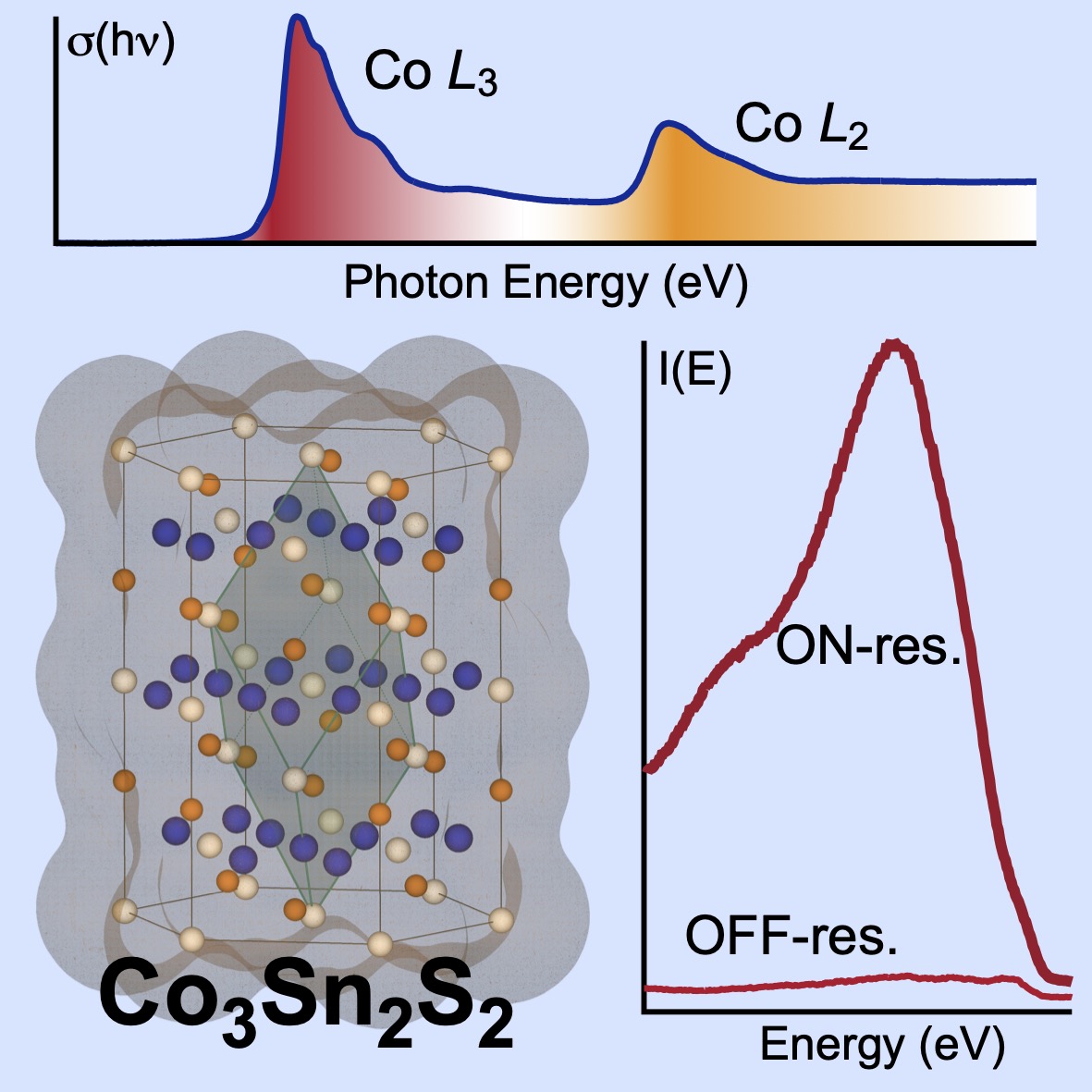}

\end{tocentry}

\begin{abstract}

The electronic structure of the natural topological semimetal Co$_3$Sn$_2$S$_2$ crystals was studied using near-edge x-ray absorption spectroscopy (NEXAFS) and resonant photoelectron spectroscopy (ResPES). Although, the significant increase of the Co\,$3d$ valence band emission is observed at the Co\,$2p$ absorption edge in the ResPES experiments, the spectral weight at these photon energies is dominated by the normal Auger contribution. This observation indicates the delocalised character of photoexcited Co\,$3d$ electrons and is supported by the first-principle calculations. Our results on the investigations of the element- and orbital-specific electronic states near the Fermi level of Co$_3$Sn$_2$S$_2$ are of importance for the comprehensive description of the electronic structure of this materials, which is significant for future applications of this material in different areas of science and technology, including catalysis and water splitting applications.

\end{abstract}


\newpage

During the past decade in condensed matter physics, chemistry and materials science a huge interest arose to the new topological phases of solids and their connection to new exciting phenomena~\cite{Hasan.2010,Bradlyn.2017,Vergniory.2019,Narang.2021}. Here, the natural topological semimetal Co$_3$Sn$_2$S$_2$, which can be considered as a quasi-two-dimensional material, attracts a lot of attention in the last years due to its interesting crystallographic structure and electronic properties~\cite{Liu.2018,Morali.2019,Yin.2019}. Initially, this material was identified as a ferromagnetic (FM) material with a Curie temperature ($T_C$) of $177$\,K~\cite{Weihrich.2004}, followed by its theoretical description as a half-metallic ferromagnet (HMF)~\cite{Dedkov.2008dhr,Holder.2009,Schnelle.2013}. Here, during the transition from the paramagnetic (PM) to the HMF state the band gap around the Fermi level ($E_F$) is formed for the spin-down channel, leaving the spin-up channel metallic. Following the systematic description of HMF materials, Co$_3$Sn$_2$S$_2$ was described as Type\,I$_{A}$ HMF~\cite{Coey.2002,Coey.2004}.

Experimental studies of bulk Co$_3$Sn$_2$S$_2$ confirmed its ferromagnetic state below $T_C$ and the measured value of magnetic moment $0.93\,\mu_B/\mathrm{f.u.}$ is very close to the theoretical value of $1\,\mu_B/\mathrm{f.u.}$ obtained within the framework of the local spin-density approximation (LSDA)~\cite{Dedkov.2008dhr,Holder.2009,Schnelle.2013}. This material was experimentally found to be strong ferromagnet with an $\langle 001\rangle$ easy magnetisation axis and the signature of the HMF state with an energy gap of $\approx300$\,meV around $E_F$ was measured in spin-resolved scanning tunnelling microscopy (STM) experiments~\cite{Jiao.2019}. Furthermore, the recent interest to the new topological phenomena in condensed matter and the extended analysis of the electronic structure of Co$_3$Sn$_2$S$_2$ identified this material as the first magnetic Weyl semimetal with broken time-reversal symmetry demonstrating giant anomalous Hall effect and topologically protected surface states~\cite{Liu.2018,Morali.2019,Yin.2019,Li.2019f0n,Liu.2019hwh}. First photoelectron spectroscopy experiments on Co$_3$Sn$_2$S$_2$, which are more than a decade old~\cite{Dedkov.2008dhr,Holder.2009}, were devoted to the identification of chemical state of elements in poly- and single-crystalline samples and comparison to the available theoretical data obtained using LSDA, and good agreement between them was found. These works also indicate the existence of surface-derived contributions in x-ray photoelectron spectra (XPS) of core levels. Further detailed angle-resolved photoelectron spectroscopy (ARPES) and STM studies revealed the existence of the topological surface states and the respective surface Fermi arcs states around the $\overline{\mathrm{K}}$ point in the electronic structure of Co$_3$Sn$_2$S$_2$(001) with different atomic terminations~\cite{Liu.2019hwh,Morali.2019}. Also the nontrivial surface states theoretically identified for PM and FM states at $0.23$\,eV above $E_F$ for all surface terminations of Co$_3$Sn$_2$S$_2$(001) were shown to be effective for the water splitting processes~\cite{Li.2019f0n}.

Despite the recent progress in the studies of this new interesting material, the full description of its electronic structure using electron spectroscopy methods is not complete, missing the detailed investigations of the element- and orbital-specific electronic states near $E_F$. Here, we present the systematic studies of the electronic structure of bulk Co$_3$Sn$_2$S$_2$ using near-edge x-ray absorption spectroscopy (NEXAFS) and resonant photoelectron spectroscopy (ResPES). These experimental methods being element specific allow to discriminate between atomic partial contributions in the valence band states around $E_F$. Our results demonstrate that the spectral weight in ResPES measurements at the Co\,$2p$ absorption edge is dominated by the strong Auger contribution. These observations indicate the delocalised character of the valence band states above $E_F$. All photoelectron spectroscopy results are accompanied by the detailed density functional theory (DFT) calculations supporting experimental findings.

\begin{figure}[h]
\centering
\includegraphics[width=0.75\textwidth]{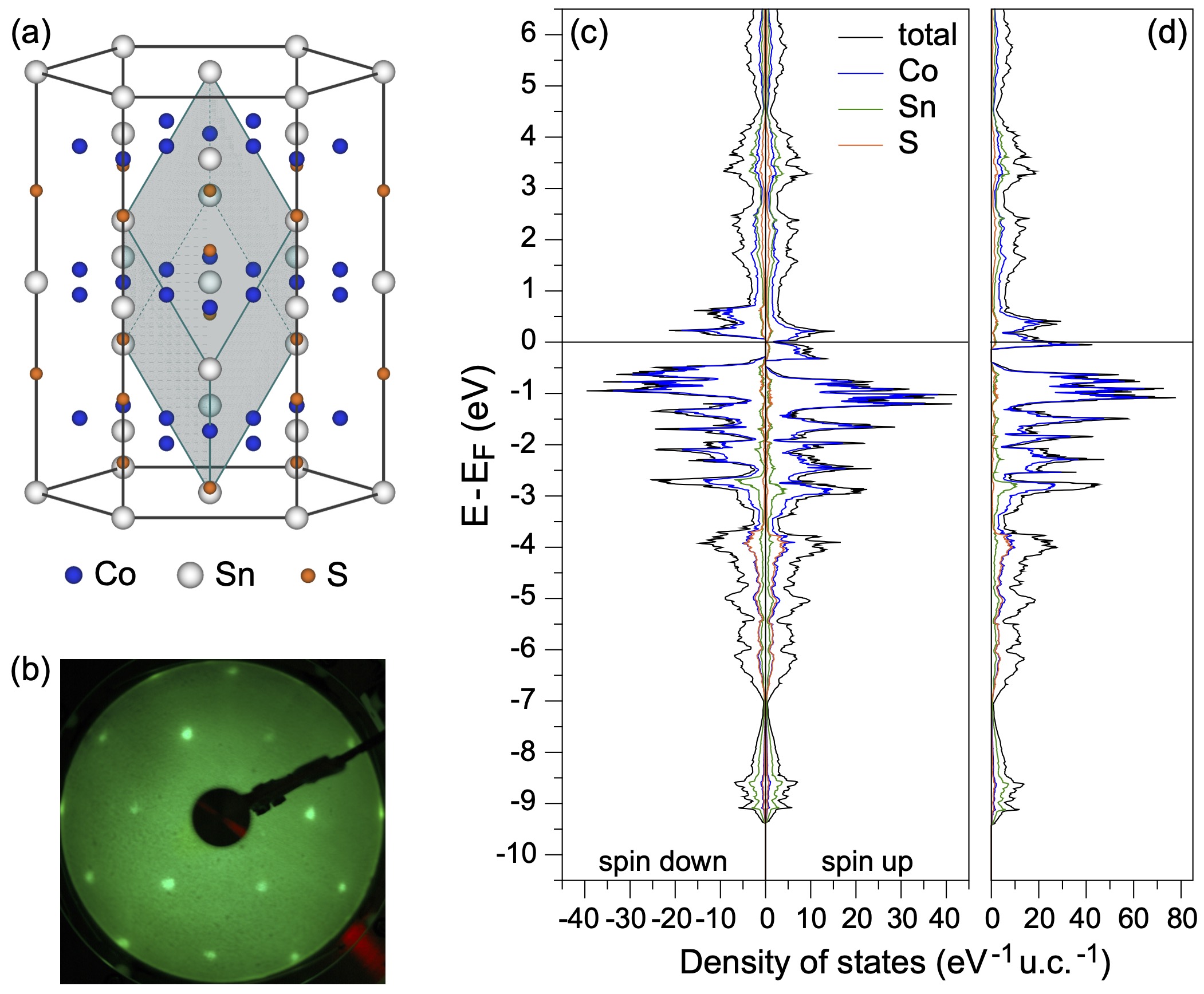}
\caption{(a) Hexagonal crystal structure of Co$_3$Sn$_2$S$_2$ with the rhombohedral primitive cell highlighted by grey colour. (b) LEED image obtained from the Co$_3$Sn$_2$S$_2$(001) surface. (c) and (d) Total and atom-projected density of states calculated for the FM and PM states of bulk Co$_3$Sn$_2$S$_2$, respectively.}
\label{fig:structure_DOS}
\end{figure}

The crystal structure of bulk Co$_3$Sn$_2$S$_2$ has a rhombohedral symmetry of the $R\overline{3}m$ space group (shandite structure isostructural to the mineral Ni$_3$Pb$_2$S$_2$, see Fig.~\ref{fig:structure_DOS}(a)). This structure can be considered and described using a hexagonal symmetry (marked in the figure) with experimentally obtained lattice parameters $a=b=5.3680$\,\AA\ and $c=13.1830$\,\AA, which are very close to the values obtained within the general gradient approximation (GGA) (5.3785\,\AA\ and 13.1594\,\AA, respectively) (see Supporting Information for the corresponding data files for the experimentally derived and theoretically optimised crystallographic structures). Here, the inclusion of the van der Waals correction in the DFT calculations leads to the slightly reduced values of lattice constants (5.3125\,\AA\ and 13.0009\,\AA, respectively) and to decreasing the magnetic moment of Co atoms ($0.347\,\mu_B$ for GGA+D3 vs. $0.353\,\mu_B$ for GGA), however, does not significantly influencing the band structure of Co$_3$Sn$_2$S$_2$. After sample was cleaved and annealed in vacuum the clear hexagonal LEED spots (Fig.~\ref{fig:structure_DOS}(b)) characteristic for the 3-fold symmetry of the Co$_3$Sn$_2$S$_2$(001) surface were obtained, indicating the correctness of the description of this material in a hexagonal structure as well as the high quality of the studied material.

Bulk Co$_3$Sn$_2$S$_2$ crystals in our study were synthesised using Sn-flux method from the stoichiometric amounts of elements and obtained several-mm$^2$ samples were characterised using different bulk- and surface-sensitive techniques. Figs.\,S1 and S2 summarises the results of structural, magnetic and transport measurements. These results confirm the high quality of the studied samples and strong ferromagnetic state of Co$_3$Sn$_2$S$_2$ with Curie temperature of $T_C=172\pm1$\,K, similar to previous results~\cite{Schnelle.2013,Liu.2018,Jiao.2019}. The GGA calculated total and atom-projected density of states (DOS) for the FM and PM states of Co$_3$Sn$_2$S$_2$ are shown in Fig.~\ref{fig:structure_DOS}(c,d), respectively and these results are in very good agreement with previous works~\cite{Dedkov.2008dhr,Schnelle.2013,Jiao.2019} (see also Fig.\,S3 for calculated band structures). The valence band states are spread down to $E-E_F\approx-10\,\mathrm{eV}$. As previously reported, in the FM state this material is HMF with only spin-up electrons at $E_F$ and the band gap of $0.340$\,eV for spin-down electrons. In the PM state the density of states at $E_F$ is significantly increased. At low binding energies (close to $E_F$) they are dominated by the Co\,$3d$ states weakly hybridised with Sn and S states. Below $E-E_F\approx-3.5\,\mathrm{eV}$ the Sn\,$5s$ and S\,$3p$ start to contribute to DOS where they effectively hybridise with Co\,$3d$ states.

\begin{figure}[h]
\centering
\includegraphics[width=0.75\textwidth]{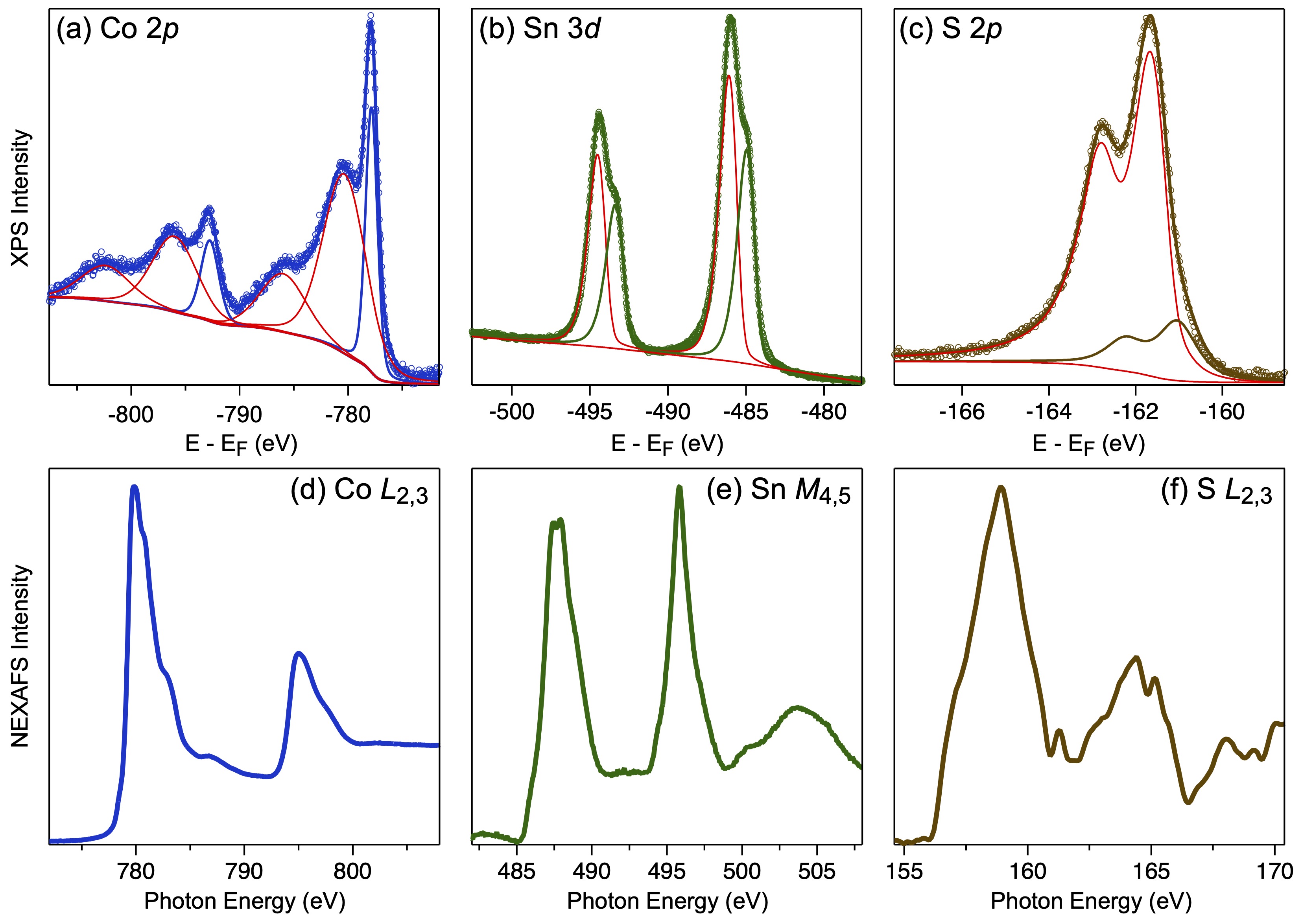}
\caption{XPS spectra of Co$_3$Sn$_2$S$_2$ collected at $h\nu = 1000$\,eV: (a) Co\,$2p$, (b) Sn\,$3d$, and (c) S\,$2p$. In case of the S\,$2p$ spectra we used spin-orbit split lines presented by red and green lines for the bulk- and surface-derived emission, respectively. NEXAFS spectra of Co$_3$Sn$_2$S$_2$ collected in the TEY mode: (d) Co\,$L_{2,3}$, (e) Sn\,$M_{4,5}$, and (f) S\,$L_{2,3}$.}
\label{fig:XPS_NEXAFS}
\end{figure}

Fig.~\ref{fig:XPS_NEXAFS} shows the representative XPS and NEXAFS spectra for Co$_3$Sn$_2$S$_2$ crystal. XPS lines of (a) Co\,$2p$, (b) Sn\,$3d$, and (c) S\,$2p$ are in very good agreement with previously published results~\cite{Holder.2009,Li.2019f0n} and confirm the respective chemical states of elements (Co$^0$)$_3$(Sn$^{2+}$)$_2$(S$^{2-}$)$_2$ (see Fig.\,S4 for the survey spectra measured after annealing of the crystal in vacuum). This is supported by the energy positions of the respective core-level lines: $E_B(\mathrm{Co}\,2p_{3/2})=777.9$\,eV corresponds to the metallic state of Co$^0$ (additional lines at $780.5$\,eV and $786.2$\,eV can be attributed to the surface oxidation and satellite structure, respectively, appearing due to the sample treatment), $E_B(\mathrm{Sn}\,3d_{5/2})=486.1$\,eV and $E_B(\mathrm{S}\,2p_{3/2})=161.7$\,eV. Both S\,$2p$ and Sn\,$3d$ spectra demonstrate the existence of emission at lower binding energies, that can be attributed to the surface-derived contributions and can be assigned to the simultaneous existence of different surface terminations for the cleaved Co$_3$Sn$_2$S$_2$(001) surface, which were observed experimentally~\cite{Li.2019f0n,Morali.2019}.

NEXAFS spectrum measured at the Co\,$L_{2,3}$ absorption edge (Fig.~\ref{fig:XPS_NEXAFS}(d)) corresponds to the metallic state of Co$^0$ with the position of the main peak at $779.8$\,eV. Additionally, two distinct shoulders at $780.6$\,eV and $782.5$\,eV on the higher energy side of the main line can be seen. As was pointed in Ref.~\cite{Li.2019f0n} the Co-atoms, which form the Kagome lattice, can be considered as placed in the strongly distorted S$_2$-Co-Sn$_4$ face-shared octahedra (see Fig.\,S5). Due to the metallic state of Co$^0$, metallic character of Co$_3$Sn$_2$S$_2$ and the relatively large distances between Co and chalcogen atoms, the splitting between $t_{2g}$ and $e_g$ is small favouring filling the $d$-states according to the Hund's rule and giving high-spin state for Co-atoms (although magnetic moments are strongly suppressed and screened by valence electrons). This is also confirmed by our DFT calculations. The existing electric field of the strongly distorted octahedra consisting of Sn and S can lead to the splitting of unoccupied Co states resulting the observed Co\,$L_{2,3}$ NEXAFS spectra. These shoulders also correlate with the positions of broad features in the unoccupied Co-projected $3d$ DOS shown in Fig.~\ref{fig:structure_DOS}(c,d). The satellite structure observed at $\approx786.8$\,eV can be assigned to changes in the hybridisation of the $sp$ orbitals between the initial and final states or a mixed ground state effect.

In case of the Sn\,$M_{4,5}$ NEXAFS spectra (Fig.~\ref{fig:XPS_NEXAFS}(e) and Fig.\,S6) the electron transitions occur from the spin-orbit split $3d_{3/2,5/2}$ core levels onto the unoccupied $p$ and $f$ shells (according to the selection rules $\Delta l=\pm1$)~\cite{Qiao.2014,Sharma.2016,Eads.2017}. The main $M_5$ peak ($487.7$\,eV) is split and additionally two intensity shoulders can be observed at the lower and higher energies at $\approx486.1$\,eV and $\approx488.7$\,eV, respectively. In Co$_3$Sn$_2$S$_2$ bulk crystal two types of Sn atoms are identified (Sn1 are located in the same plane as Co atoms and Sn2 are placed between the planes formed by S atoms). For both types of atoms their valence band orbitals effectively hybridise with the orbitals of S and Co atoms as can be deduced from the atom projected DOS in Fig.~\ref{fig:structure_DOS}(c,d). The formed hybrid bands can be seen just above $E_F$, at $E-F_F\approx2$\,eV and $E-F_F\approx3.5$\,eV. These bands with the significant Sn-atoms projected DOS can be assigned to the observed features of the Sn\,$M_{4,5}$ NEXAFS spectra. At the same time the Sn-atom projected DOS at $E-F_F\approx2$\,eV is split that can explain the observed small splitting of the Sn\,$M_5$ main line. The description and interpretation of the S\,$L_{2,3}$ NEXAFS spectra (Fig.~\ref{fig:XPS_NEXAFS}(f)) is not certain as the spin-orbit splitting of the S\,$2p$ level is small ($\approx1.1$\,eV) leading to the overlapping of the respective absorption transitions. However, the main absorption band peaked at $158.9$\,eV and the respective low- and high-energy shoulders can be assigned to the hybrid bands formed by the respective valence orbitals (Fig.~\ref{fig:structure_DOS}(c,d)).

\begin{figure}[h]
\centering
\includegraphics[width=0.75\textwidth]{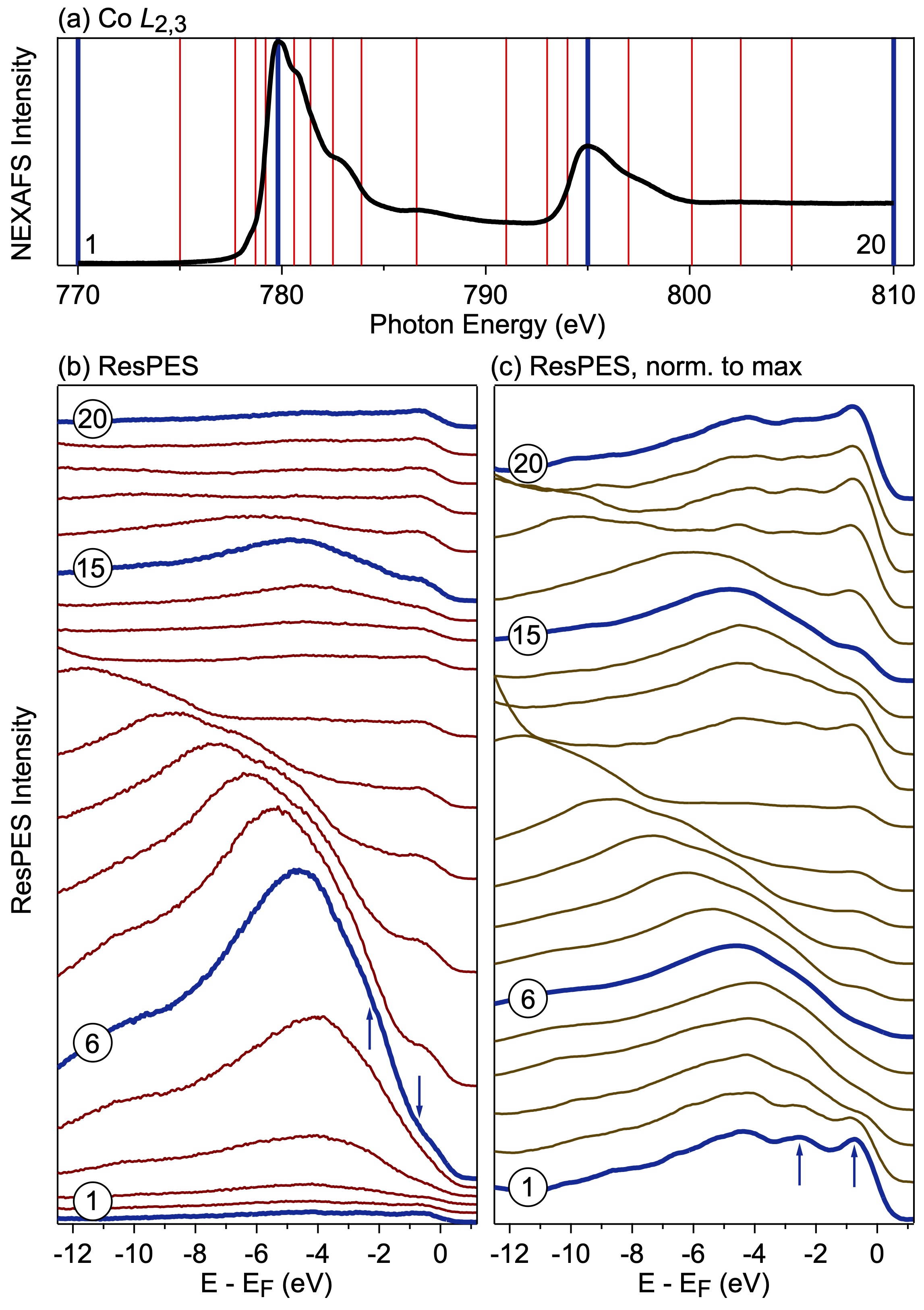}
\caption{Results of ResPES experiments for Co$_3$Sn$_2$S$_2$: (a) the reference Co\,$L_{2,3}$ NEXAFS spectrum and (b,c) a series of photoemission spectra taken at the particular photon energies marked by the corresponding vertical lines in panel (a). Spectra in panel (c) are normalised by the maximum intensity for every emission line. All ResPES spectra are shifted in the vertical direction for clarity.}
\label{fig:ResPES}
\end{figure}

Figure~\ref{fig:ResPES} summarises the result of the ResPES studies of Co$_3$Sn$_2$S$_2$ performed at the Co\,$L_{2,3}$ absorption edge: (a) Co\,$L_{2,3}$ spectrum with marked photon energies used to acquire PES spectra presented in (b) and (c), where in the later panel all spectra were normalised to the maximum intensity for every line in order to emphasise the respective spectral features. A huge increase in the photoemission intensity by factor of $\approx35$ is observed at the Co\,$L_{3}$ absorption edge (compare spectra 1 and 6). This enhancement of the Co\,$3d$ intensity, which photon energy dependence in this case is described by the Fano profile, is expected by the resonant photoelectron process described by the interference of two photoemission channels: (i) a direct photoemission $2p^63d^n + h\nu \rightarrow 2p^63d^{n-1} + e$ and (ii) a photoabsorption process followed by a participator Coster-Kronig decay $2p^63d^n + h\nu \rightarrow 2p^53d^{n+1} \rightarrow 2p^63d^{n-1} + e$, where the initial and final states are identical~\cite{Davis.1981abc}. At the same time the intensity enhancement can also occur due to the normal Auger decay described by $2p^63d^n \rightarrow 2p^53d^{n+1} \rightarrow 2p^63d^{n-2} + e$. Here, the resonance peak stays at the same binding energy, while the normal Auger line linearly shifts to higher binding energies with increasing the photon energy. The same description of the resonant processes is also valid for the spin-orbit counterpart Co\,$L_{2}$.

PES spectra of bulk Co$_3$Sn$_2$S$_2$ crystals measured with pre- and post-absorption-edge photon energies (spectrum 1 -- $h\nu=770$\,eV and spectrum 20 -- $h\nu=810$\,eV, respectively) reflect the total DOS dominated by the Co\,$3d$ states (see Fig.~\ref{fig:structure_DOS}(c,d)) and these results are in very good agreement with previous results obtained for polycrystalline samples~\cite{Dedkov.2008dhr}. At photon energies corresponding to the Co\,$L_3$ absorption edge (spectra $4-10$) a large increase of intensity is observed. The intensity maximum of the enhanced features linearly shifts to large binding energies as photon energy is increased. This behaviour indicates the normal Auger decay character of these emission lines (this consideration is valid for both spin-orbit split absorption edges). At the same time we can indicate the increase of the photoemission intensity for spectral features with no shift of the binding energy (marked by vertical arrows in Fig.~\ref{fig:ResPES}(b,c)). However, the intensity of the first feature at $E-E_F\approx=-0.75$\,eV is increased by factor of $6.5$, compared to factor $35$ for the increase of intensity associated with the normal Auger decay.

In ResPES, the resonance excitations with constant binding energy are observed when the intermediate electronic state (i.\,e., the final state for the photoabsorption) in the resonant process is relatively localised, that leads to the direct recombination of photoexcited electron. Opposite to this, for the case of normal Auger decay, if the photoexcited electron on the $3d$ orbital is delocalised, then it will relax to the other orbital before returning to the inner $2p$ shell~\cite{Zangrando.2007,Kono.2019}. Therefore the observation of the strong intensity for the normal Auger decay compared to the resonant behaviour in the ResPES spectra for Co$_3$Sn$_2$S$_2$ can be considered as a sign of the delocalisation of photoexcited electrons on the Co\,$3d$ unoccupied states. Considering the atom-projected band structure and DOS of Co$_3$Sn$_2$S$_2$ (Fig.~\ref{fig:structure_DOS}(c,d) and Fig.\,S2) one can identify two types of valence band states above $E_F$. For the PM state the flat bands in the range of $0-1$\,eV give two sharp peaks in DOS and can be responsible for the resonating behaviour of valence band features marked by arrows in Fig.~\ref{fig:ResPES}. At energies above $1$\,eV the dispersive Co\,$3d$ bands are present and the photoexcited electrons (Co\,$2p\rightarrow3d$) exhibit strong itinerancy and do not remain at the same atomic site. As shown above, such behaviour leads to the strong enhancement of the normal Auger emission. These observations clearly explain the delocalisation of electrons in Co$_3$Sn$_2$S$_2$ and reduction the effective magnetic moments of Co atoms.

\textit{In summary}, we performed systematic studies of the electronic properties of natural topological quasi-2D semimetal Co$_3$Sn$_2$S$_2$ using electron spectroscopy methods XPS, NEXAFS and ResPES. Our spectroscopic studies reveal that octahedrally coordinated, by S and Sn, Co atoms are in the high-spin state. However, the strong delocalisation of the Co\,$3d$ states and the itinerant character of valence electrons lead to the strong reduction of magnetic moment of Co atoms to $0.3\,\mu_B$. This description is strongly supported by the ResPES measurements at the Co\,$L_{2,3}$ absorption edge. The spectral weight of ResPES spectra is dominated by the normal Auger decay indicating the delocalised nature of the photoexcited Co\,$3d$ electrons. The present results give the presentation of the atom- and orbital-specific electronic states near $E_F$ and are of importance for the complete description of the electronic structure of Co$_3$Sn$_2$S$_2$ and of the localisation character of electronic states in this material. Such information is important for the future using of this material in different areas of science and technology, including catalysis and water splitting applications.

\begin{acknowledgement}
Y.~D. and E.~V. thank the ``ENSEMBLE3 - Centre of Excellence for nanophotonics, advanced materials and novel crystal growth-based technologies'' project (GA No.~MAB/2020/14) carried out within the International Research Agendas programme of the Foundation for Polish Science co-financed by the European Union under the European Regional Development Fund and the European Union’s Horizon 2020 research and innovation programme Teaming for Excellence (GA. No. 857543) for support of this work.  A.~T. acknowledges the Ministry of Science and Higher Education of the Russian Federation No. 0852-2020-0019 (State assignment in the field of scientific activity, Southern Federal University, 2020 Project No. BAZ0110/20-1-03EH). A.~M. acknowledges the BMBF (grant No.~05K19KER). D.~S. acknowledges the BMBF (grant No.~0519ODR). We thank HZB for the allocation of synchrotron radiation beamtime and for support within the bilateral Russian-German Laboratory program. The HPC Service of ZEDAT, Freie Universität Berlin, is acknowledged for computing time.
\end{acknowledgement}

\begin{suppinfo}
The following files are available free of charge
\begin{itemize}
  \item Additional theoretical and experimental data can be downloaded via link:\\https://pubs.acs.org/doi/10.1021/acs.jpclett.1c02790
\end{itemize}

\end{suppinfo}


\clearpage

\clearpage

\clearpage

\end{document}